\documentclass[a4paper,fleqn]{cas-sc}

\usepackage[numbers]{natbib}
\usepackage{subfig}
\usepackage{amsmath}
\usepackage{graphicx}
\usepackage{amssymb,amsfonts,bm}
\usepackage{algorithmic}
\usepackage{textcomp}
\usepackage{xcolor}

\def\tsc#1{\csdef{#1}{\textsc{\lowercase{#1}}\xspace}}
\tsc{WGM}
\tsc{QE}

\begin{document}
\let\WriteBookmarks\relax
\def\floatpagepagefraction{1}
\def\textpagefraction{.001}

\shorttitle{Experimental Results of Underwater SSP Inversion by Few-shot MTL}   

\shortauthors{Wei Huang et al.}

\title[mode = title]{Experimental Results of Underwater Sound Speed Profile Inversion by Few-shot Multi-task Learning}  

\tnotemark[1]

\tnotetext[1]{This document is the results of the research project funded by Natural Science Foundation of Shandong Province (ZR2023QF128), China Postdoctoral Science Foundation (2022M722990), Qingdao Postdoctoral Science Foundation (QDBSH20220202061), National Natural Science Foundation of China (NSFC:62271459), National Defense Science and Technology Innovation Special Zone Project: Marine Science and Technology Collaborative Innovation Center (22-05-CXZX-04-01-02), and the Fundamental Research Funds for the Central Universities, Ocean University of China (202313036)..}

\author[1]{Wei Huang}[orcid=0000-0002-8284-2310]
\ead{hw@ouc.edu.cn} 
\credit{Conceptualization of this study, Methodology, Software, Writing - Original draft preparation}

\author[2]{Fan Gao}
\ead{gaofan@stu.edu.cn}
\credit{Data curation}
\author[2]{Junting Wang}
\ead{wjtsci2015@163.com} 
\credit{Data curation}

\author[1]{Hao Zhang}
\ead{zhanghao@ouc.edu.cn} 
\cormark[1]

\address[1]{Ocean University of China, Qingdao 266100, China}
\address[1]{Shandong University at Weihai, Weihai 264200, China}
\cortext[1]{Corresponding author} 

\begin{abstract}
Underwater Sound Speed Profile (SSP) distribution has great influence on the propagation mode of acoustic signal, thus the fast and accurate estimation of SSP is of great importance in building underwater observation systems. The state-of-the-art SSP inversion methods include frameworks of matched field processing (MFP), compressive sensing (CS), and feedforeward neural networks (FNN), among which the FNN shows better real-time performance while maintain the same level of accuracy. However, the training of FNN needs quite a lot historical SSP samples, which is diffcult to be satisfied in many ocean areas. This situation is called few-shot learning. To tackle this issue, we propose a multi-task learning (MTL) model with partial parameter sharing among different traning tasks. By MTL, common features could be extracted, thus accelerating the learning process on given tasks, and reducing the demand for reference samples, so as to enhance the generalization ability in few-shot learning. To verify the feasibility and effectiveness of MTL, a deep--ocean experiment was held in April 2023 at the South China Sea. Results shows that MTL outperforms the state--of--the--art methods in terms of accuracy for SSP inversion, while inherits the real-time advantage of FNN during the inversion stage.
\end{abstract}


\begin{highlights}
\item To achieve good accuracy performance of SSP inversion under few-shot learning situation, we propose the MTL approach. Through learning on multi-task (different kinds of SSPs), the common features of SSPs are extracted to accelerate the convergence rate of the model on any given task with less training times, so as to weaken the over-fitting effect.
\item To verify the feasibility of MTL, a deep-ocean experiment for SSP inversion is conducted. The accuracy performance of SSP inversion is evaluated based on measured data and compared with state-of-the-art methods.
\item To solve the problem of limited coverage of XCTD, we propose a fast sound speed distribution estimation method based on MFP with EOF decomposition, which could accurately extend the SSPs to deeper layers.
\end{highlights}


\maketitle

\section{Introduction}
Underwater observation system has become a good way to provide positioning, navigation, and timing (PNT) services in recent years \cite{erol2011survey,Qu2016Surveylocalization,Luo2021Localization}. Communication and localization are the two most important technological basis for underwater observation systems, and because of the attenuation problems, the sound wave becomes the main carrier for long-distance signal transmission in underwater environment. However, the uniformly distribution of sound speed can lead to great Snell effect. which means the signal will propagate non-stragihtly. So the sound field information will be dynamically changed such as signal propagation time or received signal strength \cite{jensen2011computational}. Fortunately, if the sound speed distribution could be obtained, the sound field distribution can be accurately estimated according to ray theory \cite{munk1979ocean} or normal mode theory \cite{munk1983ocean,shang1989ocean}, thus the accuracy of ranging and positioning could be improved \cite{carroll2014demand,liu2015joint,wu2017matched}.

\indent With a same range scale, the variation of sound speed in the vertical direction is much greater than that in the horizontal direction, so sound speed profiles (SSPs) are usually used to represent the distribution of sound speed \cite{jensen2011computational}. Recently, many SSP inversion approaches leveraging sound field information have been proposed in underwater wireless sensor networks for inverting SSPs. The research of novel SSP inversion methods is very promising because they are more automatic and less labor-time-consuming than the measurement of SSPs by sound velocity profiler (SVP) or conductivity-temperature-depth (CTD) systems \cite{zhang2015inversion,huang2018underwater}.

\indent The state-of-the-art SSP inversion frameworks contains matched field processing (MFP) \cite{tolstoy1991acoustic}, compressive sensing (CS) \cite{choo2018compressive,liqian2019acoustic} and feedforward neural networks (FNN) \cite{stephan1995inverting,huang2018underwater}. The esitimation of real-time SSP is a difficulte work because, to the best of our knowledge, there is no empirical formula that could establish the mapping relationship from sound field information to the SSP. Nevertheless, based on ray theory, Tolstoy proposed a MFP framework combining empirical orthognal function (EOF) decomposition for SSP inversion. In this work, the estabilishment of direct mapping from the sound field to the sound speed distribution is avoided. Therefore, MFP has become the mainstream framework for SSP inversion for a long time.

\indent The optimal solution of \cite{tolstoy1991acoustic} is searched traversely, so the computational complexity of MFP is quite high. To accelerate the SSP inversion process, many researchers have adopted heuristic algorithms into the search for optimal solutions, such as the simulated annealing algorithm \cite{zhang2005thestudy}, the particle swarm optimization (PSO) algorithm \cite{zhang2012inversion,zheng2017improved}, and the genetic algorithm \cite{tang2006sound,sun2016inversion}. However, the core idea of heuristic algorithm is based on Monte Carlo, so the heuristic algorithm still requiters a lot of iterative searches to determine the matching items. To improve the effiency of SSP inversion, \cite{choo2018compressive,liqian2019acoustic} proposed a CS framework for SSP inversion, which establishes a dictionary to directly map the sound field information to sound speed distribution. Compared with MFP, the CS framework only requires a few iterations to train the dictionary, so that the computational complexity can be reduced. However, the mapping from sound field to sound speed distribution is linearly simplified through the first-order Taylor expansion, which sacrifices the accuracy performance.

\indent Recently, Bianco et al. \cite{Bianco2019Machine} did a comprehensive survey that machine learning has gained broad application prospects in the field of underwater acoustics, such as seafloor characterization \cite{Michalopoulou1993Application}, range estimation \cite{Komen2020Seabed}, geoacoustic inversion  \cite{Piccolo2019Geoacoustic}, and SSP inversion \cite{stephan1995inverting,huang2018underwater,Huang2021CollaboratingAI}. In our early work \cite{huang2018underwater}, we proposed a FNN structure for SSP inversion with the assist of ray theory. The model training can be completed offline in advance. After the model converging, only one round of forward propagation is required to invert the SSP when feeding the measured sound field data into the network. 
Thus, the inversion efficiency of the FNN model is significantly higher than that of the MFP and CS models. To improve the robostness of FNN models under noise interference, we proposed an auto-encoder feature-mapping neural network (AEFMNN) in \cite{Huang2021CollaboratingAI}. By denoising and reconstructing sound field information, hidden features with stronger anti-interference ability are extracted, thereby improving the accuracy of SSP inversion.

\indent The performance of MFP, CS, or FNN models all relies on a large amount of historical SSPs as references. However, due to the high economic and labor costs of historical SSPs measured by CTD or SVP, the reference SSPs available in many spatio-temporal ocean areas are so limited that there may be insufficient reference data for model training, making the model prone to be overfitting. Many approaches for few-shot learning surveyed in \cite{joaquin2018metasurvey,timothy2020metalearning} have been proposed to solve the overfitting problem, such as regularization \cite{Goodfellow2016DL}, training dataset expanding with generative adversarial networks \cite{Jin2020DL}, multi-task learning (MTL) \cite{Rich1997Multitask,Yang2017Deep}, transfer learning (TL) \cite{weiss2016survey,pan2010transfersurvey}, and meta-learning approaches \cite{joaquin2018metasurvey,timothy2020metalearning,Finn2017MetaLearning}. 

\indent Regularization establishes a way to limit the model scale by narrowing down the values of weight parameters (L2 norm) or making the model parameters sparse (L1 norm). Thus, the ability of fitting complex relationships of the model is weaken so that overfitting problem could be reduced. Training dataset expanding aims to enrich the training dataset that could represent the whole situation of target domain, however, if the original training data are unable to uniformly represent the regional distribution of soudn speed, the expanded training dataset cannot compensate for this shortcoming, thus the model is still prone to be overfitting. MTL jointly learns several related tasks, while TL uses past experience of a source task to improve learning on a new task by transferring the model's prior parameter in \cite{Hang2017Unsupervised} or the feature extractor from the solution of a previous task in \cite{Yosinski2014How}. However, the generalization ability on new tasks of these two models is still insufficient due to excessive retention of old task information.

\indent For solving the few-shot learning problem in SSP inversion, we propose a multi-task learning (MTL) method to accurately estimate the regional SSP distribution. The core idea is to extract the common features from different kinds of SSP clusters via a partial parameter shared neural network, which forms a set of initialization parameters for the model of inversion task. When training on the few-shot samples of the task, the learning rate is dynamically adjusted based on the distance of spatio-temporal imformation between the reference SSP sample and the task mission. Through MTL, the convergence rate of the model could be accelerated and the sensibility to the input data could be retained, so that the model will not be over trained on few-shot task samples. To demonstrate the effectiveness of MTL, we conducted a deep-ocean experiment in the South China Sea in April 2023. The contribution of our paper can be concluded as:

\begin{itemize}
	\item To achieve good accuracy performance of SSP inversion under few-shot learning situation, we propose the MTL approach. Through learning on multi-task (different kinds of SSPs), the common features of SSPs are extracted to accelerate the convergence rate of the model on any given task with less training times, so as to weaken the over-fitting effect.
	\item To verify the feasibility of MTL, a deep-ocean experiment for SSP inversion is conducted. The accuracy performance of SSP inversion is evaluated based on measured data and compared with state-of-the-art methods.
	\item To solve the problem of limited coverage of XCTD, we propose a fast sound speed distribution estimation method based on MFP with EOF decomposition, which could accurately extend the SSPs to deeper layers. 
\end{itemize}

\indent The rest of this paper is organized as follows. In Sec. 2, we propose the MTL model for SSP inversion. In Sec. 3, we introduce the scenarios and processes of the deep--ocean experiment for SSP inversion. Experimental results are discussed in Sec. 4, and conclusions are given in Sec. 5.

\section{MTL Model for SSP Inversion}
\begin{figure}[htbp]
	\centering
	\includegraphics[width=0.6\linewidth]{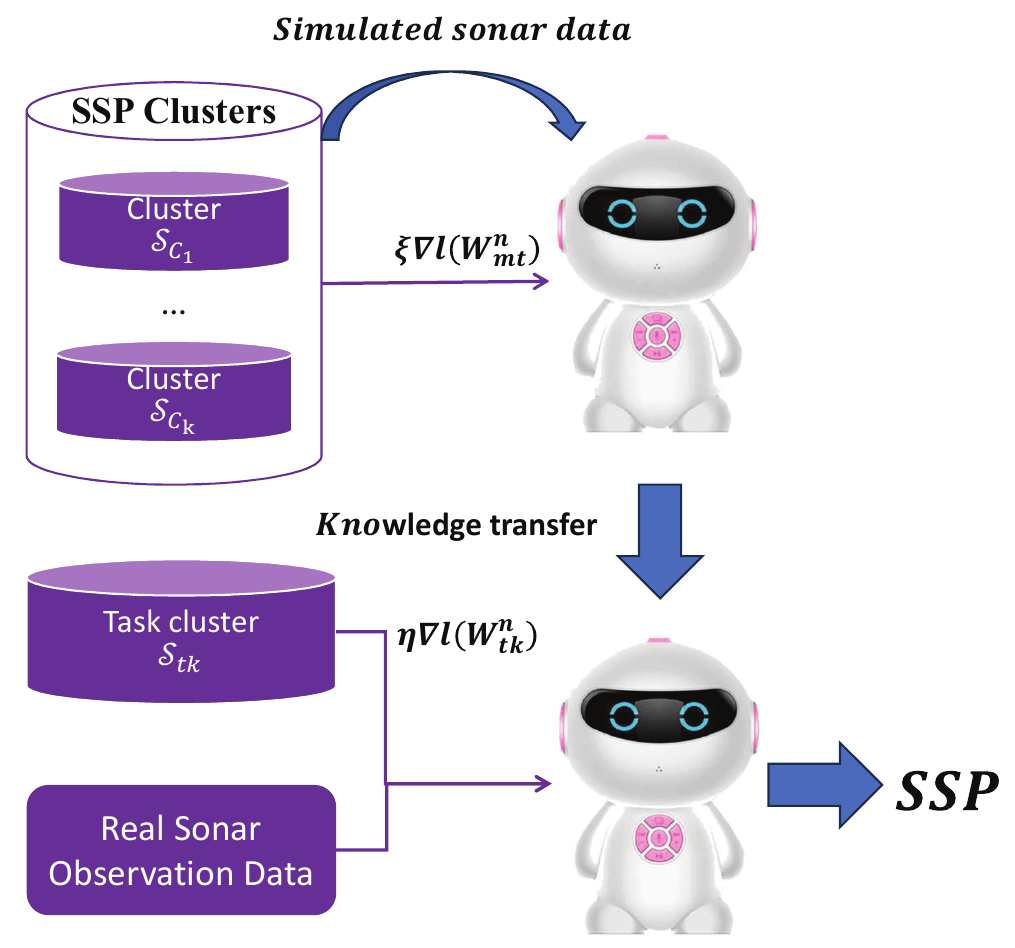}
	\caption{MTL SSP inversion model.}
	\label{fig01}
\end{figure}

\indent To improve the SSP inversion accuracy under few-shot learning situation, we propose a MTL model, in which different types of SSPs sampled in different spatio-temporal regions will be used for pretraining. While for training on the task group, the learning rate will be dynamically adjusted according to the distance of spatio-temporal information between the reference SSP and the inversion task. The illustration of MTL training and SSP inversion is shown in Fig.\ref{fig01}.

\begin{figure}[htbp]
	\centering
	\includegraphics[width=0.8\linewidth]{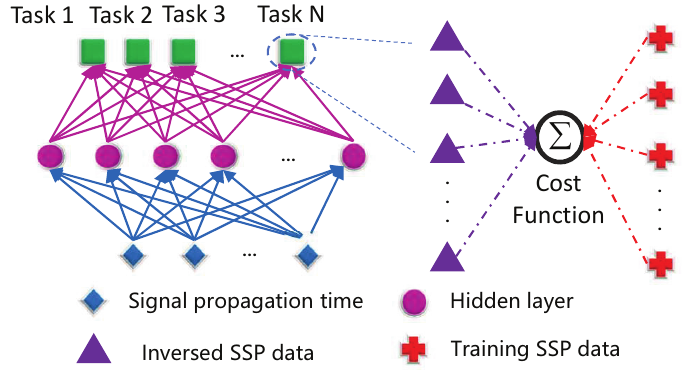}
	\caption{Structure of neural networks in MTL.}
	\label{fig02}
\end{figure}

\indent The MTL model is a three-layers FNN proposed in our previous work \cite{huang2018underwater} as shown in Fig.\ref{fig02}. There is a multi-task learner and a task learner in out model. The input layer is signal propagation time sequence, and the number of input layer neurons is consistent with the sonar observation data collected by the transmitting and receiving nodes, and the detailed sonar observation data collection process will be given in the ocean experiment section. The output layer is the inverted SSP, while the label data is the training SSP. For pretraining, the weights between the input layer and the hidden layer are shared, while the weights between the hidden layer and the output layer are unique related to the traning task. During each epoch, $n$ kinds of SSPs with V-shot SSPs for each are randomly chosen from the whole SSP clusters for training the mutli-task model, which aims to learn a good set of initialization parameters for the model's shared parameters. Thus, the task learner initialized by the mutli-task learner could be quickly converged with a few training times on the task SSP training set.

\subsection{Label SSP Data Preparation}
\indent The training historical SSP data for the mutli-task learner could be either clusterd manually or intelligently. In our pervious work \cite{Huang2023SSPCluster}, we propose a local density clustering method based on Euclidean distance, which can effectively group SSPs with similar distribution even through they are sampled in different spatio-temporal ocean regions. 

\indent For a given SSP inversion task, a small number of reference SSPs having the similar distribution with the target task are still necessary for training the task learner. However, the sound speed distribution of the target task is not a prior information, therefore, the potential reference SSPs can not be selected according to the distribution features of SSPs. 
In our pervious work \cite{Huang2023SSPCluster}, we propose a method to find proper traning data for a given task according to its spario-temporal information.

\indent In this paper, a spatio-temporal distance parameter $\phi$ is designed to describe the similarity between the sampling region of a reference SSP and the task, which can be expressed as:
\begin{equation}
	\phi = \lambda_{tk} * \phi_\alpha + (1-\lambda_{tk}) * \phi_\beta, \label{eq1}
\end{equation}
where $\phi_\alpha$ is the time difference, $\phi_\beta$ is the location difference, and $0\leq\lambda_{tk}\leq1$ is a factor to balance $\phi_\alpha$ and $\phi_\beta$. The time difference $\phi_\alpha$ is calculated by:
\begin{equation}
	{\phi _\alpha } = \left\{ {\begin{array}{*{20}{l}}
			{\left| {{\alpha _{tk}} - {\alpha _{r}}} \right|,if\left| {{\alpha _{tk}} - {\alpha _{r}}} \right| < 183}\\
			{365 + \min ({\alpha _{tk}},{\alpha _{r}}) - \max ({\alpha _{tk}},{\alpha _{r}}),otherwise}
	\end{array}} \right. \label{eq2}
\end{equation}
where $\alpha_{tk}$ and $\alpha_r$ are the time information of SSP inversion task and a random reference SSP of the task learner, respectively. The resolution of sampling time information is defined in days, because SSPs sampled at the same period in different years have almost the same distribution. Therefore, the time information is cyclically encoded that can be described as equation \eqref{eq2}.

\indent The space information is defined by the latitude and longitude coordinate of an SSP. The location difference $\phi_\beta$ is calculated by:
\begin{equation}
	{\phi _\beta } = \sqrt {{{\left( {\beta _{tk}^x - \beta _{r}^x} \right)}^2} + {{\left( {\beta _{tk}^y - \beta _{r}^y} \right)}^2}}, \label{eq3}
\end{equation}
where $\beta_{tk}^x$ and $\beta_{tk}^y$ represent the longitude and latitude coordinates of task location after coding, respectively, and $\beta_{r}^x$ and $\beta_{r}^y$ represent the longitude and latitude coordinates of SSP sampling space, respectively. For the Northern Hemisphere, the coded $\beta^y$ equals to the SSP's latitude coordinate, while $\beta^x$ is:
\begin{equation}
	\beta ^x = \left\{ {\begin{array}{*{20}{l}}
			{\left| {\hat \beta ^x} \right| - 180, if\quad 0^\circ E < \hat \beta ^x < 180^\circ E}\\
			{180 - \left| {\hat \beta ^x} \right|, if\quad 0^\circ W < \hat \beta ^x < 180^\circ W}
		\end{array},} \right. \label{eq4}
\end{equation}
where ${\hat \beta ^x}$ is the original longitude coordinate of the SSP.

\indent Through comparing the spatio-temporal distance between all historical SSPs and the target SSP, the cluster which contains most of the $\psi$ nearest SSPs will be determined as the reference of the task learner.

\subsection{Simulation of Signal Propagation Times}
\indent Both the training of multi-task learner and the task learner require sound field information as model's input under a given SSP distribution. However, the real measured sound field data are usually collected at the inversion stage, and those sound field data used for model training need to be simulated.

\indent For ocean sensing networks, underwater anchor nodes can be located by a long baseline positioning system composed of several buoys, while the coordinates of buoys can be obtained via the global positioning system (GPS). Let a buoy node be the signal sender, and $M$ underwater anchor nodes be the receivers, the horizontal propagation distance of the signal with a given SSP $S=[(s^1,1),...,(s^d,d)]$ can be calculated according to our previous derivation in \cite{Huang2021CollaboratingAI} as:

\begin{equation}
	\begin{aligned}
		h^m &=\frac{s^1}{\cos\theta^{1,m}}\sum _{d=1}^{D-1}\left | \frac{\Delta z_{d}}{s^{d+1}-s^{d}} \left( \sqrt{\Gamma^m_{d}}-\sqrt{\Gamma^m_{d+1}} \right) \right |,\\
		\Gamma^m_{d} &= 1 - \left(\frac{\cos\theta^{1,m}}{s^1} \right)^2(s^{d})^2,\label{eq5}
	\end{aligned}
\end{equation}
where $h^m$ is the horizontal signal propagation distance from the buoy to the $m$th anchor node, $d$ is the index of depth layers, $\theta^{1,m}$ is the initial grazing angle from source to the $m$th receiver, $s^d$ is the sound speed value at depth with index $d$ and $\Delta z_{d}$ is the depth difference of the linear SSP at the $d$th layer with depth boundaries of $d$ and $d+1$.

\indent With known horizontal distance $h^m$, the grazing angle of signal from source to each receiver can be searched by equation \eqref{eq5}. Then the ideal signal propagation time from source to the $m$th receiver can be simulated as:
\begin{equation}
	t^m = \sum\limits_{d = 1}^{D - 1} {\left| {\frac{{\Delta {z_{d}}}}{{s^{d + 1} - s^{d}}}\ln \left( {\frac{{s^{d }\left( {1 + \sqrt {{\Gamma ^m_{d+1}}} } \right)}}{{s^{d + 1}\left( {1 + \sqrt {{\Gamma ^m_{d}}} } \right)}}} \right)} \right|},\label{eq6}\\
\end{equation}
It can be noticed that $t^m$ is also a function of the initial grazing angle $\theta^{1,m}$.

\subsection{Training of the Multi-task Learner}
\indent The first training phase of MTL that can be conducted offline is the training of multi-task learner. In this stage, the neuron connection parameter of multi-task learner is randomly initialized as $W_{mt,h}^1$ and $W_{mt,o}^1$. $W_{mt,h}^1$ is the weight matrix between the input layer and the hidden layer, and $W_{mt,o}^1$ is the weight matrix between the hidden layer and the output layer.

\indent At the beginning of the $p$th epoch \footnote{An epoch indicates a round of parameter updating.} (or iteration), $N$ SSP clusters are randomly chosen from the $K$ available training SSP clusters ($N \leq K$) for training the multi-task learner. In each selected SSP cluster, there are $V$ training ($S_v, v=1,2,...,V$). For SSP cluster $n$, the $V$ SSPs are used for one step learning with the cost function defined as:
\begin{equation}
	{l^{(n)}}\left( {W_{mt,n}^p} \right) = \sum\limits_{v = 1}^V \left(\frac{1}{2}{\sum\limits_{d = 1}^D {{{\left( {s_v^d - \tilde s_v^d} \right)}^2}} }+{\left\| W_{mt,n}^p \right\|_1}\right), W_{mt,n}^p = \left[W_{mt,h,n}^p,W_{mt,o,n}^p\right], \label{eq7}
\end{equation}
where $s_v^d$ is the sound speed of the $v$th training SSP at depth with index of $d$, $\tilde s_v^d$ is the corresponding inverted sound speed, and ${\left\|  W_{mt,n}^p \right\|_1}$ is the regularization item. Next, the local parameters are updated by back propagation (BP) algorithm \cite{Rumelhart1986Learning}:
\begin{equation}
	\dot{W} _{mt,n}^p = W _{mt,n}^p - \frac{\xi}{N} {\nabla _{W _{mt,n}^p}}{l^{(n)}}\left( {W _{mt,n}^p} \right), \label{eq8}
\end{equation}
where $\xi$ is the learning rate of the multi-task learner. 

\indent For a specified SSP inversion task, the parameters of multi-task learner $\dot W_{mt,h}^P$ is transferred as the initialization for the task learner, thus $W _{tk,h}^1 = \dot W_{mt,h}^P$, where $W _{tk,h}^1$ is the weight value matrix between the input layer and the hidden layer. The weight value matrix $W _{tk,o}^1$ between the hidden layer and the output layer will be initialized randomly. The task learner will be trained on a few reference SSPs by one or a few steps to form the converged model $\dot W _{tk}$. 

\indent Let the $i$th reference SSP be  $S_{tk,ri}=\left[\left(s_{tk,ri}^1,1\right),...,\left(s_{tk,ri}^d,d\right)\right]$, $d=1,2,...,D$ with coded sampling location as $\left(\beta _{tk,ri}^x,\beta _{tk,ri}^y\right)$ and time information as $\alpha _{tk,ri}$, so the spatio-temporal distance $\phi_{tk,ri}$ between the $i$th task reference SSP and the inversion task (with coded location as $\left(\beta _{tk}^x,\beta _{tk}^y\right)$ and time information as ($\alpha _{tk}$) can be calculated according to equation \eqref{eq1}, \eqref{eq2} and \eqref{eq3}.

\indent To reduce the negative transfer effect of MTL, we propose a dynamic learning rate adjustment strategy based on the inverse spatio-temporal distance weighted coefficient for task learner training. For the $i$th training SSP, the learning rate of task learner $\eta _{tk,ri}$ will be:

\begin{equation}
	\eta _{tk,ri} = \xi\frac{\frac{1}{\phi_{tk,ri}}}{\sum\limits_{i = 1}^I \frac{1}{\phi_{tk,ri}}}.  \label{eq9}
\end{equation}
where $\phi_{tk,ri}$ is calculated by \eqref{eq3} with $\lambda_{tk,ri} = 0.9$.

\indent If there are total $J$ epochs for task learner training, the cost function $l_{tk}\left( {W _{tk}^j} \right)$ of the $j$th ($j=1,2,...,J$) batch will be:
\begin{equation}
	l_{tk}\left( {W _{tk}^j} \right) = \frac{1}{2}\sum\limits_{d = 1}^D {{{\left( {s_{tk,ri}^d - \tilde s_{tk,ri}^d} \right)}^2}},  \label{eq10}
\end{equation}
where $\tilde s_{tk,ri}^d$ is the inverted SSP related to reference SSP $S_{tk,ri}$. Then, the parameter updating of task learner is conducted according to:
\begin{equation}
	\dot{W} _{tk}^j = W _{tk}^j - \eta _{tk,ri} {\nabla _{W _{tk}^j}}{l_{tk}}\left( {W _{tk}^j} \right). \label{eq11}
\end{equation}
At last, the converged model will be $\dot W _{tk} = \dot{W} _{tk}^J$.

\indent When feeding measured sound field information into model $\dot W _{tk}$, the inverted SSP $\widetilde S_{tk}$ can be estimated via once forward propagation.

\section{Deep--ocean Experiments}
\subsection{Experimental Settings}
\indent To evaluate the feasibility and effectiveness of proposed MTL method for SSP inversion under few-shot learning situation. We conducted deep--ocean experiments at the South Sea of China with areas of $10 km \times 10 km$ in middle April 2023, where the depth is about 3500 meters. The relevant data collection corresponding to SSP inversion lasted for a total of 3 days. 

\begin{figure}[htbp]
	\centering
	\includegraphics[width=0.6\linewidth]{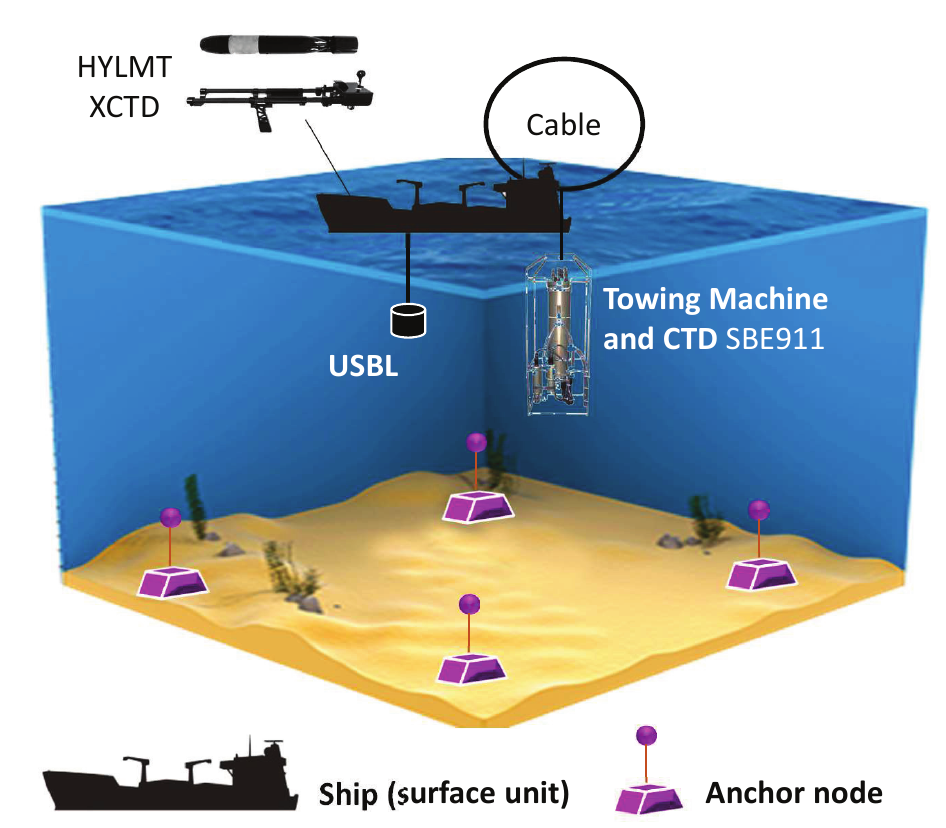}
	\caption{System composition of ocean experiments.}
	\label{fig03}
\end{figure}

\indent The system composition is shown in Fig.\ref{fig03}, including 4 anchor nodes, a ship unit that containing a CTD, a set of expendable CTD (XCTD), and an ultra short baseline (USBL) system fixed to the right side of our ship. The anchor nodes and USBL system were used for collecting sonar observation data, while SSP samples were collected by CTD and XCTD.

\begin{figure}[htbp]
	\centering
	\includegraphics[width=0.75\linewidth]{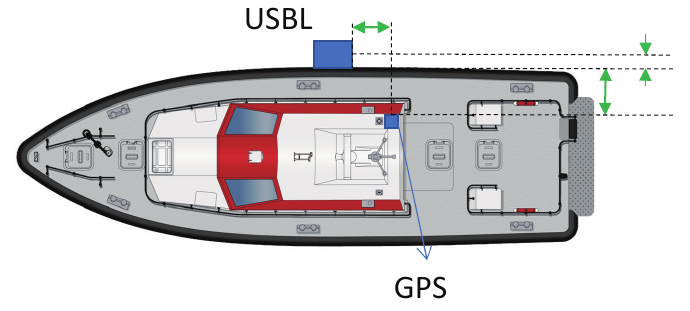}
	\caption{Relative azimuth of GPS and USBL in horizontal direction.}
	\label{fig04}
\end{figure}

\indent At the beginning, 4 anchor nodes were sunk in turns to the seabed and their positions were calibrated using signal round-trip propagation time measured by USBL in a circular trajectory. These 4 anchor nodes formed a diamond topology. The real--time position of USBL was located through a ship borne GPS, which was installed near the central axis of the ship. In order to improve the position accuracy of anchor nodes, the lever--arm error between USBL and GPS needs to be corrected. The distance measurement of lever--arm in horizontal direction between the GPS and USBL is shown in Fig.\ref{fig04}. The horizontal distance was measured in three sections, and the vertical distance was about $18.86$ meters ($7.5$ meters under the water surface) measured under a basically wave free environment in a harbor. Let GPS be the coordinate origin, then the relative coordinates of USBL would be $\left(6.774,8.392,-18.8603\right)$ in meters.

\begin{figure}[htbp]
	\centering
	\includegraphics[width=0.5\linewidth]{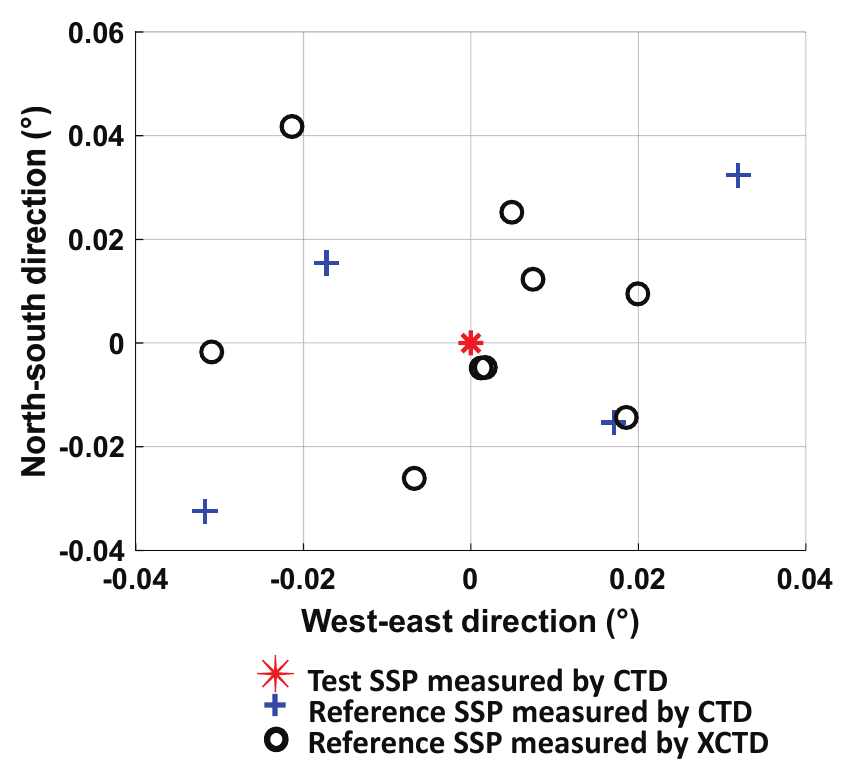}
	\caption{Space distribution of sampled SSPs.}
	\label{fig05}
\end{figure}

\indent For deploying each anchor node, a full depth of SSP was measured through ship borne CTD, the product model of which is SBE911 produced by Sea-bird Scientific \cite{SEABIRD}. Considering the high time costs of SSP measurement by CTD (almost 3 hours for once measurement with no ship movement), the XCTDs were adopted to collect the other 9 SSPs, the model of which is HYLMT-2000 produced by \cite{HAIYAN}. XCTD provides a fast way for SSP measurement that can be performed during ship navigation, and the time cost is related to the measurement depth. For HYLMT-2000 used in this experiment, it takes only about 20 minutes to measure an SSP with maximum depth of 2000 meters. The CTD and XCTD were arranged at the stern of the ship, and the water entry coordinates of CTD and XCTD were measured by the real--time ship borne GPS. These 13 SSPs were collected as reference SSPs. For testing, the last full depth of SSP was measured as an SSP inversion task at the topology center among the 4 anchor nodes. At the same time, the USBL interacted with four anchor nodes to collect sonar observation data at a period of 8 seconds. The location of anchor nodes and sampled SSPs are shown in Fig.\ref{fig05}.

\subsection{MFP-based SSP Extending}
\indent Although XCTD has obvious advantages of time efficiency compared with the SBE911 CTD, the depth coverage of XCTD is limited due to the pressure resistance characteristics of sensors. To tackle this issue, we propose a fast sound speed distribution estimation method based on MFP with EOF decomposition, which is briefly called EOF–PSSP–ME. 

\begin{figure}[htbp]
	\centering
	\includegraphics[width=0.7\linewidth]{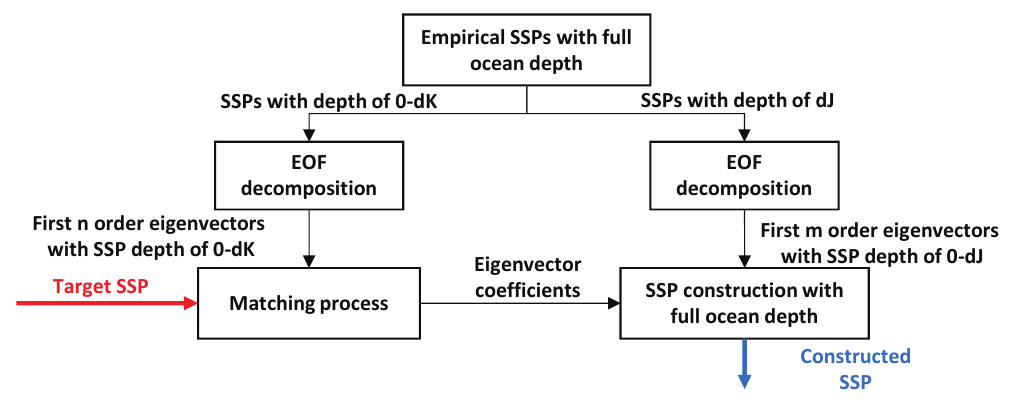}
	\caption{The workflow of EOF--PSSP--ME.}
	\label{fig06}
\end{figure}

\indent The workflow is shown in Fig. \ref{fig06}. There are four steps in EOF--PSSP--ME. Firstly, the original empirical SSPs with full ocean depth are intercepted, forming SSP data with partial depth that is the same as target SSP (to be extended). Secondly, the empirical SSPs with partial and full ocean depth are both decomposed by EOF. Then, the eigenvector coefficients of the target SSP on eigenvectors of empirical SSPs with partial ocean depth are calculated through matching process. Finally, the target SSP with full ocean depth could be constructed by combining the eigenvector coefficients with the eigenvectors of original empirical SSPs. To simplify the model introduction without lose of generality, the SSPs are assumed to have been interpolated at intervals of 1 meter depth.

\paragraph{Empirical SSPs Interception}
\indent Let empirical SSPs with full ocean depth form SSP data set $\bm{\mathcal{S}}=\left\{\bm{S_{1}},\ldots,\bm{S_{i}}\right\}$, each SSP sample could be expressed as:
\begin{equation}\label{eq12}
	\bm{S_{i}}=\left\{ {\left( {s_{i,0},d_{0}} \right),\left( {s_{i,1},d_{1}} \right),\ldots,\left( {s_{i,j},d_{j}} \right)} \right\},
\end{equation}
where $s_{i,j}$ is sound speed, $d_{j}$ is depth, $i=1,2,\ldots,I,$ means the $i$th SSP sample, $j=0,1,\ldots,J$ is the index label of depth, and the maximum common depth of SSPs with full ocean depth is $d_{J}$.

\indent If the maximum depth of target SSP to be extended is $d_{K}$, all SSPs in $\bm{\mathcal{S}}$ will be partially intercepted by depth and form a data set of reference SSPs $\bm{\mathcal{\bar{S}}}=\left\{\bm{\bar{S_{1}}},\ldots,\bm{\bar{S_{i}}}\right\}, i=1,2,\ldots,I$ with maximum depth that equals to $d_{K}$. $\bm{\bar{S_{i}}}$ can be expressed as:

\begin{equation}\label{eq13}
	\bm{\bar{S_{i}}}=\left\{ {\left( {\bar{s}_{i,0},d_{0}} \right),\left( {\bar{s}_{i,1},d_{1}} \right),\ldots,\left( {\bar{s}_{i,k},d_{k}} \right)} \right\},
\end{equation}
where $k=0,1,\ldots,K$.

\paragraph{EOF Decomposition}
To maintain the original principal component features of any target SSP to be extended, it is necessary to first extract the distribution features of reference SSPs $\bm{\mathcal{\bar{S}}}$, and obtain the components of the target SSP on these distribution features.

\indent  The average SSP distribution of $\bm{\mathcal{\bar{S}}}$ is:
\begin{equation}\label{eq14}
	\bm{\bar{S}_{ar,d_{K}}} = \left\{ {\left( {{\bar{s}}_{ar,0},d_{0}} \right),\left( {{\bar{s}}_{ar,1},d_{1}} \right),\ldots,\left( {{\bar{s}}_{ar,k},d_{k}} \right)} \right\},
\end{equation}
where ${\bar{s}}_{ar,k},k=0,1,\ldots,K$ is the average sound speed at depth $d_{k}$ that calculated by:
\begin{equation}\label{eq15}
	{\bar{s}}_{ar,k} = \frac{1}{I}\left( \bar{s}_{1,k} + \bar{s}_{2,k} + \ldots + \bar{s}_{I,k}\right).
\end{equation}

Let $\bm{S_{i}^s}$ represents the sound speed vector of $\bm{S_{i}}$ that $\bm{S_{i}^s}= \left[s_{i,0},s_{i,1},\ldots,s_{i,J}\right]$. Similarly, $\bm{\bar{S_{i}^s}} = \left[\bar{s}_{i,0},\bar{s}_{i,1},\ldots,\bar{s}_{i,K}\right]$ and $\bm{\bar{S}_{ar,d_{K}}^s} = \left[{\bar{s}}_{ar,0},{\bar{s}}_{ar,1},\ldots,{\bar{s}}_{ar,K}\right]$, where $K \leq J$. 
Then a residual matrix $\bm{S_{X,d_{K}}}$ of sound speed less than depth $d_{K}$ could be constructed as:
\begin{equation}\label{eq16}
	\bm{S_{X,d_K}} = \left\lbrack \bm{\bar{S}_1^s} - \bm{\bar{S}_{ar,d_K}^s},\bm{\bar{S_2^s}} - \bm{\bar{S}_{ar,d_K}^s},\ldots,\bm{\bar{S_I^s}} - \bm{\bar{S}_{ar,d_K}^s} \right\rbrack.
\end{equation}

\indent Based on $\bm{S_{X,d_{K}}}$, the covariance matrix $\bm{C_{S,d_{K}}}$ could be derived as:
\begin{equation}\label{eq17}
	\bm{C_{S,d_{K}}} = \frac{1}{I}\bm{S_{X,d_{K}}} \times \bm{{S_{X,d_{K}}}^{T}},
\end{equation}
where $\bm{C_{S,d_{K}}}$ is a matrix with orders of $(K+1) \times (K+1)$. 

\indent Through EOF decomposition, the matrix of feature values $\bm{\Lambda_{d_{K}}}$ and matrix of feature vectors $\bm{V_{d_{K}}}$ could be obtained, which satisfies:
\begin{equation}\label{eq18}
	\bm{C_{S,d_{K}}} \times \bm{V_{d_{K}}} = \bm{V_{d_{K}}} \times \bm{\Lambda_{d_{K}}},
\end{equation}
In \eqref{eq18}, $\bm{V_{d_{K}}}=\left[\bm{v_{1,d_K}},\bm{v_{2,d_K}},\ldots,\bm{v_{n,d_K}}\right],n=1,2,\ldots,N$ is a matrix with orders of $(K+1) \times N$, and each column represents a feature vector. $\bm{\Lambda_{d_{K}}}$ is an N-order diagonal matrix that can be expressed as:
\begin{equation}\label{eq19}
	\bm{\Lambda_{d_{K}}} = \begin{bmatrix}
		\lambda_{1,d_{K}} & \cdots & 0 \\
		\vdots & \ddots & \vdots \\
		0 & \cdots & \lambda_{N,d_{K}} \\
	\end{bmatrix},
\end{equation}
where each feature value $\lambda_{n,d_{K}},n=1,2,\cdots,N$ maps to one feature vector $\bm{v_{N,d_K}}$.

\indent To estimate the full--depth distribution of the target SSP, the feature vectors and values of empirical SSPs with full ocean depth need to be extracted. The average SSP distribution of $\bm{\mathcal{S}}$ is:
\begin{equation}\label{eq20}
	\bm{S_{ar,d_J}}=\left\{{\left( s_{ar,0},d_0 \right),\left( s_{ar,1},d_1 \right),\ldots,\left( s_{ar,j},d_j \right)} \right\},
\end{equation}
where $s_{ar,j},j=0,1,\ldots,J$ is the average sound speed at depth $d_{j}$ that calculated by:
\begin{equation}\label{eq21}
	s_{ar,j} = \frac{1}{I}\left( s_{1,j} + s_{2,j} + {\ldots + s}_{I,j}\right).
\end{equation}
Similarly to $\bm{S_{X,d_{K}}}$, a residual matrix $\bm{S_{X,d_{J}}}$ of sound speed up to depth $d_{J}$ could be constructed as:
\begin{equation}\label{eq22}
	\bm{S_{X,d_J}} = \left\lbrack \bm{S_1^s} - \bm{S_{ar,d_J}^s},\bm{S_2^s} - \bm{S_{ar,d_J}^s},\ldots,\bm{S_I^s} - \bm{S_{ar,d_J}^s} \right\rbrack,
\end{equation}
where $\bm{S_{ar,d_J}^s} = \left[s_{ar,0},s_{ar,1},\ldots,s_{ar,J}\right]$.
The covariance matrix $\bm{C_{S,d_{J}}}$ could be derived as:
\begin{equation}\label{eq23}
	\bm{C_{S,d_{J}}} = \frac{1}{I}\bm{S_{X,d_{J}}} \times \bm{{S_{X,d_{J}}}^{T}},
\end{equation}
where $\bm{C_{S,d_{J}}}$ is a matrix with orders of $(J+1) \times (J+1)$.

Refer to \eqref{eq18}, the matrix of feature values $\bm{\Lambda_{d_{J}}}$ and vectors $\bm{V_{d_{J}}}$ according to empirical SSPs with full ocean depth satisfy:
\begin{equation}\label{eq24}
	\bm{C_{S,d_{J}}} \times \bm{V_{d_{J}}} = \bm{V_{d_{J}}} \times \bm{\Lambda_{d_{J}}},
\end{equation}
$\bm{V_{d_{J}}}=\left[\bm{v_{1,d_J}},\bm{v_{2,d_J}},\ldots,\bm{v_{m,d_J}}\right],m=1,2,\ldots,M$ is a matrix with orders of $(J+1) \times M$. $\bm{\Lambda_{d_{J}}}$ is an M-order diagonal matrix that can be expressed as:
\begin{equation}\label{eq25}
	\bm{\Lambda_{d_{J}}} = \begin{bmatrix}
		\lambda_{1,d_{J}} & \cdots & 0 \\
		\vdots & \ddots & \vdots \\
		0 & \cdots & \lambda_{M,d_{J}} \\
	\end{bmatrix}.
\end{equation}

\indent It has been summarized that the difference in SSPs could be represented by combining feature vectors of 3--6 orders. Therefore, $\bm{V_{d_{J}}}$ and $\bm{V_{d_{K}}}$ are sorted based on the corresponding feature values in descending order, and the first $\bar{M}$ or $\bar{N}$ order's feature vectors are remained, which are denoted as $\bm{\widetilde{V}_{d_J}}=\left[\bm{\tilde{v}_{1,d_J}},\bm{\tilde{v}_{2,d_J}},\ldots,\bm{\tilde{v}_{m,d_J}}\right],m=1,2,\ldots,\bar{M}$ and $\bm{\widetilde{V}_{d_K}}=\left[\bm{\tilde{v}_{1,d_K}},\bm{\tilde{v}_{2,d_K}},\ldots,\bm{\tilde{v}_{n,d_K}}\right],n=1,2,\ldots,\bar{N}$, where $3 \le \bar{M},\bar{N} \le 6$.

\paragraph{Matching Process}
To ensure that the shallow part of reconstructed SSP is consistent with the target SSP, a matching process needs to be performed. Specifically, it is a matching of the target SSP and the combination of feature vectors $\bm{\widetilde{V}_{d_{K}}}$. By projecting the target SSP onto $\bm{\widetilde{V}_{d_{K}}}$, the projection coefficient ${\bm{cf_{d_K}}}$, which is the combination coefficient of feature vectors, could be obtained. Conversely, the target sample can be restored by combining the coefficient ${\bm{cf_{d_K}}}$ with $\bm{\widetilde{V}_{d_{K}}}$. 

\indent Let the target SSP be:
\begin{equation}\label{eq26}
	\bm{S_t}=\left\{\left(s_{t,0},d_0\right),\left(s_{t,1},d_1\right),\ldots,\left(s_{t,k},d_k\right)\right\},
\end{equation}
where $k=0,1,\ldots,K$. Then the sound speed vector can be briefly expressed as $\bm{S_t^s}=\left[s_{t,0},s_{t,1},\ldots,s_{t,K}\right]$, and the residual vector will be $\bm{S_{X,t}} = \left\lbrack \bm{S_t^s} - \bm{\bar{S}_{ar,d_K}^s} \right\rbrack$.

\indent Through matching process, the coefficient $\bm{cf_{d_K}}$ could be solved by:
\begin{equation}\label{eq27}
	\bm{cf_{d_K}}=\bm{\widetilde{V}_{d_K}^T} \bullet \bm{S_{X,t}}.
\end{equation}

\paragraph{SSP Construction}
When combining $\bm{cf_{d_K}}$ and $\bm{\widetilde{V}_{d_J}}$, the target SSP with full ocean depth will be constructed:
\begin{equation}\label{eq28}
	\bm{\hat{S}_t^s}=\bm{S_{ar,d_J}^s}+\bm{\widetilde{V}_{d_J}}\bullet \bm{cf_{d_K}},
\end{equation}
where $\bm{\hat{S}_t^s}$ will be $\bm{\hat{S}_t^s}=\left[\hat{s}_{t,0},\hat{s}_{t,1},\ldots,\hat{s}_{t,J}\right]$. Considering depth information, the estimated target SSP can also be expressed as:
\begin{equation}\label{eq29}
	\bm{\hat{S}_t} = \left\{\left(\hat{s}_{t,0},d_0\right),\left(\hat{s}_{t,1},d_1\right),\ldots,\left(\hat{s}_{t,j},d_j\right)\right\},
\end{equation}
where $j=0,1,\ldots,J$.

\section{Results and Discussions}
\subsection{Preprocessing of SSP Data}
For data-driven SSP inversion methods such as EOF--MFP and deep learning, the amount of experimental data collected is relatively small, so the model is prone to be over--fitting. To improve the accuracy of SSP estimation in few-shot learning situations, we propose the MTL model and dynamically adjusting the learning rate during task learner training stage.

\indent As mentioned above, the SSP measured at the center of 4 anchor nodes is used for testing, while the other 13 SSPs are used for task learner training. However, there is still a lack of reference data for multi-task learners to extract common features of sound speed distribution. To solve this problem, 300 historical SSPs (covering at least 3500 meters depth) sampled from the Pacific, Atlantic, and Indian Oceans of the last 10 years are adopted, which are clustered into 10 groups. These SSP data come from the world ocean database 2018 (WOD'18) \cite{WOD2018}.

\begin{figure}[!htbp]
	\begin{minipage}[b]{0.45\textwidth}
		\centering
		\subfloat[]{\includegraphics[width=0.9\textwidth]{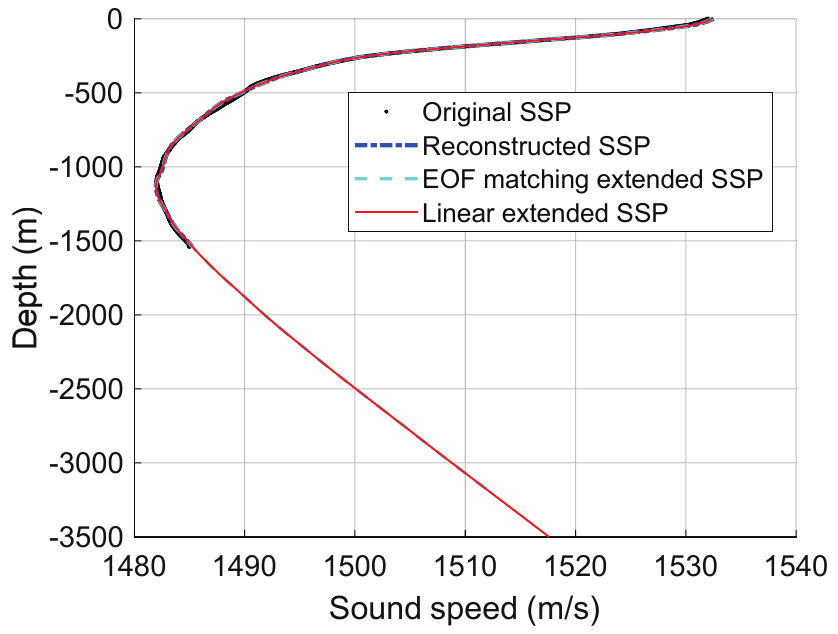}}
		\label{fig7a}
	\end{minipage}
	\begin{minipage}[b]{0.45\textwidth}
		\centering
		\subfloat[]{\includegraphics[width=0.9\textwidth]{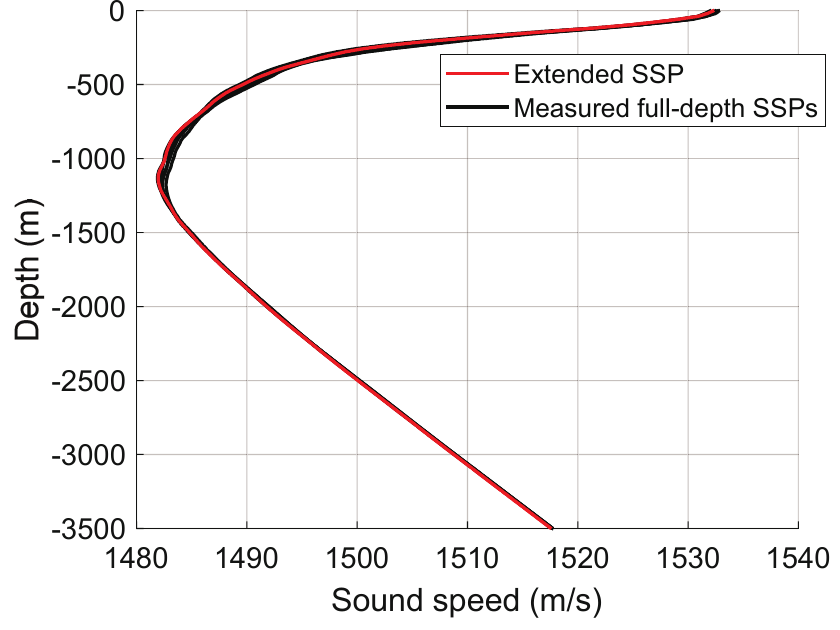}}
		\label{fig7b}
	\end{minipage}
	\caption{SSP extending. \textbf{(a)} An example of SSP extending. \textbf{(b)} An extended SSP and measured full--depth SSPs.}
	\label{fig07}
\end{figure}

\indent For most SSPs especially those sampled by XCTD, the depth scale can not cover the propagation depth of sonar signal, so the SSPs need to be extended. In this paper, SSP samples are extended through two steps, namely EOF matching extension for SSPs less than 3200 meters depth and linear extension for SSPs more than 3200 meters depth but less than 3500 meters depth. The core idea of EOF matching extension is to maintain the measured results and extend the unmeasured depth portion according to the regional empirical SSP distribution. In EOF matching extension, the full--depth and partial--depth principal components of the empirical SSPs are first extracted separately, and then the projection coefficients are obtained by projecting the measured partial--depth SSP onto the partial--depth principal components, and finally the full--depth SSP is composed of the projection coefficients and the full--depth principal components. Detailed algorithm of EOF matching extension can be referred to our work in \cite{Huang2023Fast}. For linear extention, For linear extension, we set the gradient computation window to be 50 meters, and the SSP is linearly extended based onthe gradient of the last 50 meters of the measured SSP up to the specified depth. An example of EOF matching extending and linear extending is given in Fig.\ref{fig07}.

\indent During the preprocessing of SSPs, SSPs are all standardized and interpolated with a depth spacing of 1 meter, and are finally evenly divided into 50 depth layers to form down--sampled data for being reference data of neural networks.

\subsection{Performance of MTL}
To evaluate the accuracy performance of MTL, the root mean squared error (RMSE) of inverted SSP at the center location of 4 anchors is compared with the real measured SSP. The parameter settings of SL--ML is given in Table~\ref{tab1}, and data are processed by 'Matlab 2023a'.

\begin{table}[!htbp]
	\caption{Parameter Settings of SL--ML}
	\label{tab1}
	\begin{tabular}{c|c}
		\toprule
		Training SSP clusters & 10 \\
		\hline
		SSP Clusters per epoch $K$ & 3 \\
		\hline
		SSPs for multi-task learner training (per cluster) $S_v$ & 10\\
		\hline
		Multi-task learner training epochs * & 20\\
		\hline
		Task learner training epochs & 20\\
		\hline
		Task training SSPs per epoch & 5\\
		\hline
		Maximum SSP depth & 3500 m \\
		\hline
		Points of simplified SSPs & 50\\
		\hline
		Learning rate $xi$ & 0.000002 \\
		\hline
		Task learning rate & 0.01\\
		\hline
		Input layer neurons & 120\\
		\hline
		Hidden layer neurons & 300\\
		\hline
		Output layer neurons & 50\\
		\hline
		Factor for task classification $\lambda_{tk}$ & 0.02\\ 
		\hline
		Factor for learning rate adjustment $\lambda_{tk,ri}$ & 0.9\\ 
		\bottomrule
	\end{tabular}
	\newline{\footnotesize{* One epoch corresponds to a round of parameter updating, using 3 SSP clusters.}}
\end{table}

\indent For further comparison of accuracy performance, three state--of--the--art baseline methods are also adopted: spatial interpolation (SIP), EOF-MFP and FNN. For SIP, the proportional weight follows the principle of inverse proportion of spatial distance. For EOF-MFP, the PSO is utilized for searching matching coefficients with 20 particles and 30 iterations, and the order of the principal component is 3. While for FNN, the network structure is the same as the task learner (or any base learner) of MTL, and the learning rate is $0.01$, which is equal to that of task learner in MTL.

\indent Table~\ref{tab2} gives the average accuracy performance of inverted SSPs with 100 repeated results. It shows that the accuracy of MTL outperforms other state--pf--the--art methods under few-shot learning situation, implying that the mapping relationship from sound field data to SSP distribution can be better and fasted captured through MTL. An example of inverted SSP is shown in Fig.\ref{fig07}.
\begin{table}[!htbp]
	\caption{RMSE of Inverted SSP by Different Methods}
	\label{tab2}
	\begin{tabular}{c|c|c|c|c}
		\toprule
		Methods & SIP (m/s) & EOF-MFP (m/s) & FNN (m/s) & \textbf{MTL} (m/s)\\
		\hline
		Average RMSE & 0.3895 & 0.3341 & 0.2653 & \textbf{0.2113}\\
		\hline
		0-200 (m) & 0.6875 & 0.5578 & 0.3366 & \textbf{0.1560}\\
		\hline
		200-800 (m) & 0.7529 & 0.6675 & 0.2086 & \textbf{0.1930}\\
		\hline
		800-1300 (m) & 0.4144 & 0.3743 & 0.3232 & \textbf{0.1291}\\
		\hline
		1300-3500 (m) & 0.0694 & 0.0617 & 0.2567 & \textbf{0.1998}\\
		\bottomrule
	\end{tabular}
\end{table}

\begin{figure}[htbp]
	\centering
	\includegraphics[width=0.5\linewidth]{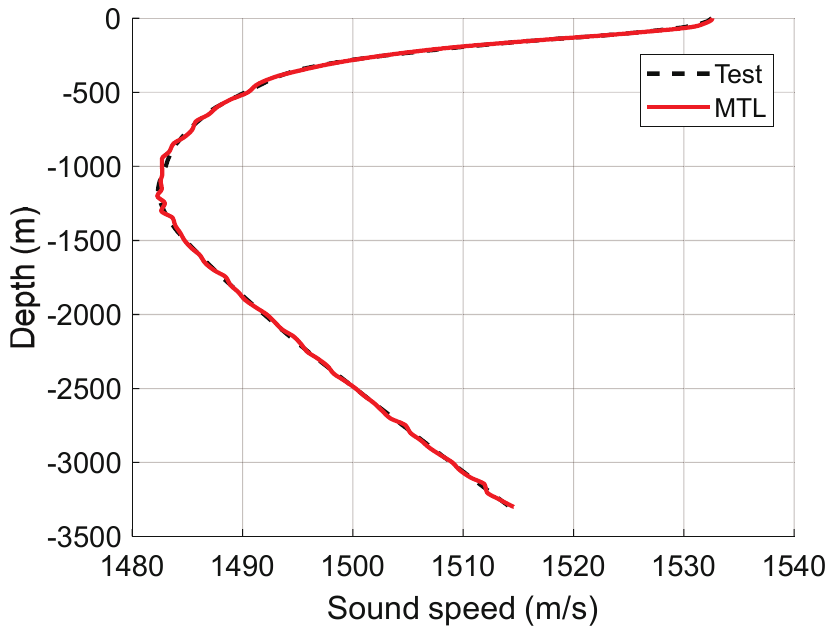}
	\caption{An example of inverted SSP.}
	\label{fig08}
\end{figure}

\indent To illustrate the reason for the fast convergence of MTL, the convergence performance of MTL is compared with that of FNN in Fig.\ref{fig08}. After learning different types of SSPs during meta learning stage, the initialization parameters of the task learner in the MTL are closer to the converged parameters, which is beneficial for faster convergence. Moreover, due to prior knowledge of sound speed distribution, the model learning process has less fluctuations, which is conducive to reaching the convergence state faster.

\begin{figure}[htbp]
	\centering
	\includegraphics[width=0.5\linewidth]{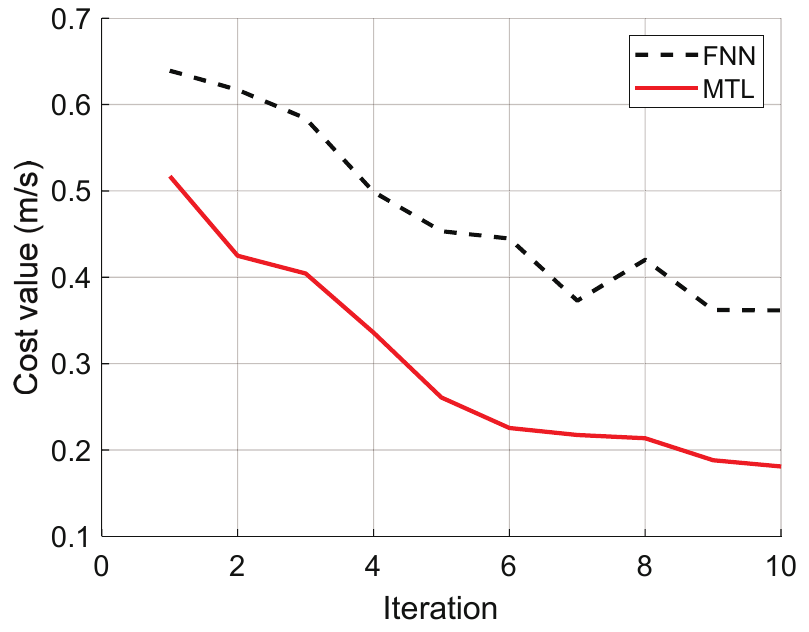}
	\caption{Convergence performance of MTL and FNN.}
	\label{fig09}
\end{figure}

\indent Since the training of neural networks for SSP inversion can be done offline, the time consumption during SSP inversion stage are more noteworthy. The average time consumption of different methods are given in Table~\ref{tab3}. MTL inherits the time efficiency advantages of FNN during the inversion stage because it only needs once forward propagation when feeding signal propagation time into the model, and this process can be done by matrix operation.

\begin{table}[!htbp]
	\caption{Time Efficiency of Inverted SSP by Different Methods}
	\label{tab3}
	\begin{tabular}{c|c|c|c|c}
		\toprule
		Methods & SIP & EOF-MFP & FNN & \textbf{MTL}\\
		\hline
		Inversion stage (s) & 0.0033 & 38.1980 & 0.0005 & 0.0008 \\
		\bottomrule
	\end{tabular}
\end{table}

\section{Conclusion}
In this paper, we propose an MTL method for SSP inversion to improve the accuracy under few-shot learning situations. For verifying the feasibility and effectiveness of the proposed model, a deep-ocean experiment at the South China Sea was conducted in April 2023. Through verification on real sampled SSP data and sonar observation data, it is shown that the proposed MTL has better accuracy performance in few-shot learning SSP construction issues, and inherits the high time efficiency of neural networks during the inversion stage.




\section*{Conflict of Interest Statement}
The authors declare that the research was conducted in the absence of any commercial or financial relationships that could be construed as a potential conflict of interest.
\bibliographystyle{cas-model2-names}

\bibliography{cas-sc-template}



\end{document}